\documentclass[mathleft
]{an}
\usepackage{graphicx}
\usepackage{amsmath}
\usepackage{times}
\begin{document}

% The following seven commands are intended for editorial usage and should be ignored by
% the author(s).
\Pagespan{0}{}% Document's page range. 
% If second parameter is left empty, the last page is computed automatically.
\Yearpublication{2011}%
\Yearsubmission{2011}%
\Month{}%   
\Volume{}%  
\Issue{}% 
% \DOI{This.is/not.aDOI}% 

\title{Spectral classification of Pleiades brown dwarf candidates\thanks{Based on observations made with ESO telescopes at the La Silla Paranal Observatory under programme ID 084.C-0654 }\\}

\author{M. Seeliger\thanks{\email{martin.seeliger@uni-jena.de}}
%Example 
%for footnote, note the usage of the \texttt{fnmsep}
%command as separator between institute number and footnote mark} 
\and  R. Neuh\"auser \and T. Eisenbeiss
}
%\titlerunning{Instructions for authors}
%\authorrunning{T.H.E. Editor \& G.H. Ostwriter}
\institute{
Astrophysikalisches Institut und Universit\"ats-Sternwarte Jena, 
Schillerg\"asschen 2-3, 07745 Jena, Germany
}

\received{23 Aug 2011}
\accepted{25 Sep 2011}
%\publonline{later}


\abstract{%
\emph{Aim: } We report on the results of the spectroscopy of 10 objects previously classified as brown dwarf candidates via RIJHK colors by Eisenbeiss et al. (\cite{Eisenbeiss2009}), who performed deep imaging observations on a $\sim$0.4~sq.deg. field at the edge of the Pleiades. We describe and judge on classification techniques in the region of M-type stars.\newline
\emph{Methods: } To classify and characterise the objects, visual and near infrared spectra have been obtained with VLT FORS and ISAAC. The spectral classification was performed using the shape of the spectra as well as spectral indices that are sensitive to the spectral type and luminosity class of M-type stars and late M-type brown dwarfs. Furthermore a spectrophotometric distance was calculated and compared the distance of the Pleiades to investigate the membership probability. As a second argument we  analyzed the proper motion.\newline
\emph{Results: } The brown dwarf candidates were found not to be brown dwarfs, but late-K to mid-M-type dwarf stars. Based on the obtained distance and tabulated proper motions we conclude that all objects are background dwarf stars.
  }

\maketitle

\section{Introduction: Our sample}

In 2007 and 2008 a $\sim$0.4~sq.deg. field ($\alpha=3^h42^m20.6^s$, $\delta=+25^\circ{}36'54''$) at the edge of the Pleiades was observed at the University Observatory Jena in order to find variable stars and brown dwarfs. First results from the variability study in this field are pubilshed in Moualla et al. (\cite{Moualla2011}) with evidence for a new early-M type Pleiades flare star. A comparison of the brown dwarf space density at the center of the Pleiades and its edge (e.g. our field) and the proper motions of brown dwarfs at its edge might show whether brown dwarfs often get ejected from the cluster center.

As a result of a deep imaging analysis, Eisenbeiss et al. (\cite{Eisenbeiss2009}) reported on the discovery of seven brown dwarf candidates and three objects close to the border to brown dwarfs being set to the spectral type M6. This border is set by the high-mass brown dwarf Teide 2 (Mart\'in et al. \cite{Martin1998}). Since the classification was based on photometric analysis, R \& I from the University of Jena observatory plus JHK from 2MASS, the spectral type of these 10 objects had to be confirmed spectroscopically. In this region of the spectral sequence, the most dominant spectral features come from TiO an VO absorption in the optical part of the spectrum. These bands are highly sensitive to the spectral type. To distinguish between giants and dwarf stars, gravity sensitive sodium and potassium indices in the optical regime are commonly used, e.g. see Hamilton \& Stauffer (\cite{Hamilton1993}) and Steele \& Jameson (\cite{Steele1995}). In near infrared spectra, the relation between the CO absorption bands and the Na doublets is sensitive to the luminosity class.

The observations are described in section~\ref{sec_observations}. The methods used to determine spectral type and luminosity class are specified in section~\ref{sec_methods}. Finally we present the results of the analysis in section~\ref{sec_results}.

\section{Observations and Data Reduction}\label{sec_observations}

The spectra presented in this work were obtained using the FOcal Reducer and low dispersion Spectrograph (FORS) and the Infrared Spectrometer And Array Camera (ISAAC), both at ESO VLT. The FORS spectra were obtained using grism GRIS\_600RI+19 in LSS mode with a wavelength coverage from $\lambda=5120\,$\AA{} to $\lambda=8450\,$\AA{} and a resolution of 1000 at the central wavelength $\lambda=6780\,$\AA{}. The ISAAC H band spectra were obtained using a slit width of 1\,arcsec and the SH filter with a wavelength coverage from $\lambda=1.40\,\mu$m to $\lambda=1.82\,\mu$m and a resolution of 500 at the central wavelength $\lambda=1.65\,\mu$m. The ISAAC K band spectra were obtained also using a slit width of 1\,arcsec and the SK filter with a wavelength coverage from $\lambda=1.82\,$nm to $\lambda=2.50\,\mu$m and a resolution of 450 at the central wavelength $\lambda=2.20\,\mu$m. The observations are summarized in Table~\ref{tab_observations}. In addition to each scientific spectrum, wavelength calibrations were done and standard stars with the same airmass were observed. The reduction was carried out using the data reduction pipeline provided by ESO for the Data File Organiser \textit{Gasgano}\footnote{http://www.eso.org/sci/software/gasgano/} (Izzo et al. \cite{Izzo2004}). This pipeline automatically reduces the input data regarding wavelength calibrations, spatial curvature calibrations, flat field distortions, sky subtraction, object detection and extraction, flux normalization and adding up of all frames obtained. The relative flux calibration was done with \textit{IRAF}\footnote{http://iraf.noao.edu/} (Tody \cite{Tody1993})
using the standard stars observed for each set of frames, and the Stellar Spectral Flux Library (Pickles \cite{Pickles1998}) as template catalogue. The relative flux correction was performed regarding the shape of the spectrum, telluric features have not been corrected. Finally, all spectra have been corrected with an extinction of $A_V=0.75$ as an average extinction for this region of the sky (Eisenbeiss et al. \cite{Eisenbeiss2009}) using the \textit{deredden}-procedure of \textit{IRAF}.

\begin{table*}
	\centering
	\caption{The observations carried out with FORS and ISAAC, the latter one in H band and K band. The I band magnitudes given in column two are taken from Eisenbeiss et al. (\cite{Eisenbeiss2009}). The number of images made with each instrument (\# images) and the exposure time per image (DIT) are given in the respective columns.}
	\label{tab_observations}
	\begin{tabular}{lcrrrrrr}
		\hline
		2MASS-number      &I$_{\textnormal{Eisenbeiss}}$& \multicolumn{2}{c}{FORS} & \multicolumn{2}{c}{ISAAC--H} & \multicolumn{2}{c}{ISAAC--K} \\
		                  & [mag]          & \# images  &  DIT [s]    &  \# images   & DIT [s]       & \# images   & DIT [s]          \\
		\hline
		J03433088+2531443 & 18.78$\pm$0.19 &  1~\textbar~ 3 & 92~\textbar~303& 42    & 120           & 100         &  90 \\
		J03430237+2530225 & 17.31$\pm$0.17 &  3         &  64         & 4            &  90           & 8           &  90 \\
		J03430027+2522082 & 18.79$\pm$0.22 &  3         & 343         & 18           & 120           & 54          &  90 \\
		J03425334+2523044 & 19.28$\pm$0.25 &  6         & 630         & 36           & 120           & 84          &  90 \\
		J03423828+2543104 & 18.07$\pm$0.18 &  3         & 151         & 8            & 120           & 14          &  90 \\
		J03423655+2542193 & 18.13$\pm$0.18 &  3         & 151         & 8            & 120           & 18          &  90 \\
		J03421030+2529316 & 18.76$\pm$0.20 &  3         & 235         & 54           &  90           & 54          &  90 \\
		J03414296+2540432 & 19.16$\pm$0.23 &  7         & 512         & 14           & 120           & 22          &  90 \\
		J03414281+2528328 & 18.12$\pm$0.18 &  3         & 112         & 8            & 120           & 16          &  90 \\
		J03413516+2546444 & 18.32$\pm$0.18 &  6         & 236         & 16           & 120           & 54          &  90 \\
		\hline
	\end{tabular}
\end{table*}

\section{Spectral classification methods}\label{sec_methods}

The spectral classification was performed using spectral indices and continuum fits. Since the continuum of a spectrum suffers from interstellar extinction, the use of indices offers the possibility to neglect the effect of extinction. The spectral index $R_{\mathrm{ind}}$ is defined by 
\begin{equation}
	\label{eqn_index}
	R_{\textnormal{ind}}=\frac{F_{W}}{F_{\textnormal{cont}}}
\end{equation}
where $F_{W}$ and $F_{\mathrm{cont}}$ are the flux of the spectral feature and the nearby contiuum, respectively.
In addition, the equivalent width of gravity, age, and metallicity sensitive absorption lines is used to determine the luminosity class and evolutionary state.

\subsection{Continuum fits}

One way to determine the spectral type of a single star is to compare the shape of the spectrum to already classified template spectra. In the range of late K-type to early L-type dwarf stars and brown dwarfs, there are three major catalogues that provide such spectra in the wavelength range of FORS: the Stellar Spectral Flux Library (Pickles \cite{Pickles1998}), the Keck LRIS spectra of late-M, L and T dwarfs\footnote{http://stsci.edu/$\sim$inr/ultracool.html} (the spectra taken for this work are from Kirkpatrick et al. \cite{Kirkpatrick1999}), and the M Dwarf and Giant Spectral Standards\footnote{Courtesy of Kelle Cruz; www.astro.caltech.edu/$\sim$kelle/M\_standards/}. In the infrared regime, the IRTF Spectral Library (Rayner et al. \cite{Rayner2009} and Cushing et al. \cite{Cushing2005}) provides a good coverage. We fitted these template spectra by eye and by using a code from Sebastian \& Guenther (\cite{Sebastian2011}). This code was operated directly on the observed spectra as it takes extinction into account, while we used extinction corrected spectra for by-eye inspection.

\subsection{Spectral indices and equivalent widths}\label{sec:spectralindices}

Following the spectral sequence down to the cool K and M stars, the atmospheres are getting cold enough for molecules to become stable. In late K-type stars TiO becomes stable first, soon followed by VO. These stabilities result in absorption features. Since the strength of absorption lines and bands depend on various aspects, spectral indices were found to be not only useful for determining the spectral type, but also for getting constraints on the luminosity class. Basic investigations of the relation of TiO and VO to the spectral type, luminosity class and absolute magnitude have been made by Cruz \& Reid (\cite{Cruz2002}). They used spectral indices from Reid, Hawley \& Gizis (\cite{Reid1995}), Kirkpatrick et. al (\cite{Kirkpatrick1999}) and Mart\'in et al. (\cite{Martin1999}). In addition to TiO and VO, CaH and CaOH also turned out to be useful for classification. An overview of the indices used is given in Table~\ref{tab_CruzReid}. 

\begin{table}[htb]
	\centering
	\caption{The wavelength ranges used by Cruz \& Reid (\cite{Cruz2002}) for the different indices used in this study.}
	\label{tab_CruzReid}
	\begin{tabular}{ccc}
		\hline
		index &$\lambda_{\textnormal{cont}}$ [\AA{}]& $\lambda_{W}$ [\AA{}]\\
		\hline
		TiO-5 & 7042 -- 7046           & 7126 -- 7135\\
		VOa   & 7430 -- 7470           & sum of 7350 -- 7370 \& 7550 -- 7570  \\
		CaH-2  & 7042 -- 7046           & 6814 -- 6846\\
		CaOH  & 6345 -- 6354           & 6230 -- 6240\\
		\hline
	\end{tabular}

\end{table}

In the optical part of the spectrum of M-type stars, TiO is producing the most dominant absorption features, showing a steplike structure between 7020$\,$\AA{} and 7150$\,$\AA{}. The first clear signs of this absorption is visible in late-K stars. The depths, hence the index-value, increasing towards late M-type stars. Therefore the ratio between the deepest absorption at 7130$\,$\AA{} and the nearby continuum (i.e. TiO-5) exhibits a clear correlation with the spectral subtype. Cruz \& Reid (\cite{Cruz2002}) found a correlation
with an uncertainty of 0.5 subclasses for stars of spectral type M0.5 to M7. The dependency inverts after spectral type M7 and the TiO-5 index is decreasing again for very late M-type and L-type stars. The same behavior can be seen for VO. Since this molecule becomes stable at lower temperatures its influence on the spectrum is stronger for later spectral types. Cruz \& Reid (\cite{Cruz2002}) found the shape of the spectrum between 7300$\,$\AA{} and 7600$\,$\AA{}, which is dominated by VO-absorption, to be sensitive to the spectral type. 
Their correlation is valid for stars between M4 and M9 with an error of 0.8 subclasses and also inverts for stars later than M9. Since there is almost no effect due to VO to be seen for early M-type stars, the index is not usable there.

To determine the luminosity class four different properties have been investigated. (a) The position of the objects in the TiO-5--CaH-2 diagram, (b) the equivalent width of the potassium doublet at $\lambda=7655/7699\,$\AA{}, (c) the equivalent  widths of the sodium doublets at $\lambda=5890/5896\,$\AA{} and $\lambda=8183/8195\,$\AA{} and (d) the relation between the sodium and calcium absorption features to the CO absorption bands in the K band. 

As Cruz \& Reid (\cite{Cruz2002}) could show, it can be distinguished between the populations of M-type dwarfs and subdwarfs by using a TiO-5--CaH-2 diagram. Since CaH-2 is sensitive to metalicity, which is the significant distinctive feature, we employ this method to exclude our stars to be subdwarfs (or even extreme subdwarfs), but not to decide between dwarfs and giants.

As shown by Steele \& Jameson (\cite{Steele1995}), sodium (i.e. the sodium doublet at $\lambda=8183/8195\,$\AA{}) can be used as a gravity indicator, hence to determine the luminosity class. According to Hamilton \& Stauffer (\cite{Hamilton1993}) the same accounts for the sodium doublet at $\lambda=5890/5896\,$\AA{} and the potassium doublet at $\lambda=7665/7699\,$\AA{}. They report that it indicates low luminosity, if these lines are \textit{``comparatively quite strong.''}

In the infrared region the relation between the strength of the sodium doublet at $\lambda\approx2.21\,\mu$m and calcium triplet at $\lambda\approx2.26\,\mu$m to the CO absorption bands at $\lambda\approx2.28-2.38\,\mu$m also shows a gravity dependence. While the first two absorption features are equally weak in giants, especially the sodium doublet is much more prominent in dwarf stars and comparably strong regarding the nearby CO absorption bands, which are strong both in giant and dwarf stars.

\subsection{Absolute magnitude, distance and proper motion}

Cruz \& Reid (\cite{Cruz2002}) also investigated the relation between the absolute magnitude of M dwarf stars and the TiO-5, CaH and CaOH indices and found the relations for each of the values given in equations\,\ref{eqn:absmag1}--\ref{eqn:absmag6}.

{\setlength{\mathindent}{0pt}%%% 
\begin{align}
	M_J=&2.79\cdot \left(\textnormal{TiO-5}\right)^2 -7.75\cdot \left(\textnormal{TiO-5}\right)+10.49\label{eqn:absmag1}\\
	&\textnormal{TiO-5}\geq0.34,\,\sigma=0.35\,\textnormal{mag}\nonumber\\
	M_J=&-7.43\cdot \left(\textnormal{TiO-5}\right)+11.82\label{eqn:absmag2}\\
	&\textnormal{TiO-5}\leq0.43,\,\sigma=0.19\,\textnormal{mag}\nonumber\\
	M_J=&6.50\cdot \left(\textnormal{CaH-2}\right)^2 -13.24\cdot \left(\textnormal{CaH-2}\right)+12.10\label{eqn:absmag3}\\
	&\textnormal{CaH-2}\geq0.36,\,\sigma=0.33\,\textnormal{mag}\nonumber\\
	M_J=&-9.33\cdot \left(\textnormal{CaH-2}\right)+12.50\label{eqn:absmag4}\\
	&\textnormal{CaH-2}\leq0.42,\,\sigma=0.19\,\textnormal{mag}\nonumber
\end{align}
\begin{align}
	M_J=&5.67\cdot \left(\textnormal{CaOH}\right)^2 -11.99\cdot \left(\textnormal{CaOH}\right)+11.82\label{eqn:absmag5}\\
	&\textnormal{CaOH}\geq0.36,\,\sigma=0.33\,\textnormal{mag}\nonumber\\
	M_J=&-7.31\cdot \left(\textnormal{CaOH}\right)+11.80\label{eqn:absmag6}\\
	&\textnormal{CaOH}\leq0.43,\,\sigma=0.28\,\textnormal{mag}\nonumber
\end{align}}

The overhead where one index value matches two absolute magnitudes corresponds roughly to the spectral type M3.5--M4.5. Reid \& Cruz (\cite{Reid2002}) argue that this is due to degeneracy which becomes more important at M$\,\approx0.1\,$M$_{\odot}$. Absolute $M_J$ magnitudes (equations\,\ref{eqn:absmag1}--\ref{eqn:absmag6}) and photometric apparent J band magnitude by the measurements of 2MASS define the spectrophotometric distance, given by the distance modulus:
\begin{equation}
	\label{eqn:distancemodulus}
	d\left[\textnormal{pc}\right]=10\cdot 10^{0.2\cdot\left(m_J-M_J-0.282\cdot A_V\right)}
\end{equation}

$A_V$ denotes the interstellar extinction, which is assumed to be the mean extinction of this region of the sky (the transformation to the J band was applied using the relation given in Rieke \& Lebofsky \cite{Rieke1985}). Obviously there are two problems. First, the measurements of 2MASS have been made several years before the FORS-spectra. Since M stars tend to be variable the luminosity might have changed. Second, the fit of the absolute magnitude by Cruz \& Reid (\cite{Cruz2002}) shows a spread of up to $0.3\,$mag. In addition, taking an  average extinction of $A_V=0.5...1\,\textnormal{mag}$ determined by Eisenbeiss et. al (\cite{Eisenbeiss2009}) introduces an error of up to $0.5$\,mag or even more in the case of background stars. This should be considered regarding the results.

Finally we carried out a proper motion analysis. Assuming that all 10 objects are Pleiades members they should follow the Pleiades mean proper motion. Five of the objects in this study can be found in the PPMXL catalogue (Roeser et al. \cite{Roeser2010}), one in NOMAD (Zacharias et al. \cite{Zacharias2004}). To compare the position and proper motion of these objects with Pleiades member stars, the WEBDA database\footnote{http://www.univie.ac.at/webda/} (Mermilliod  \cite{Mermilliod1998}) has been used to compile a list of 752 stars with listed coordinates (Hog et al. \cite{Hog2000}, Deacon \& Hambly \cite{Deacon2004}, Stauffer et al. \cite{Stauffer1998}) and a membership probability $>$75\% according to Deacon \& Hambly (\cite{Deacon2004}) or Schilbach et al. (\cite{Schilbach1995}). Of those, 221 had proper motion measurements listed in the PPMXL catalogue and therefore can be used to perform a proper motion analysis.

\section{Results}\label{sec_results}

\begin{figure*}[htb]\centering
\includegraphics[width=0.9\textwidth]{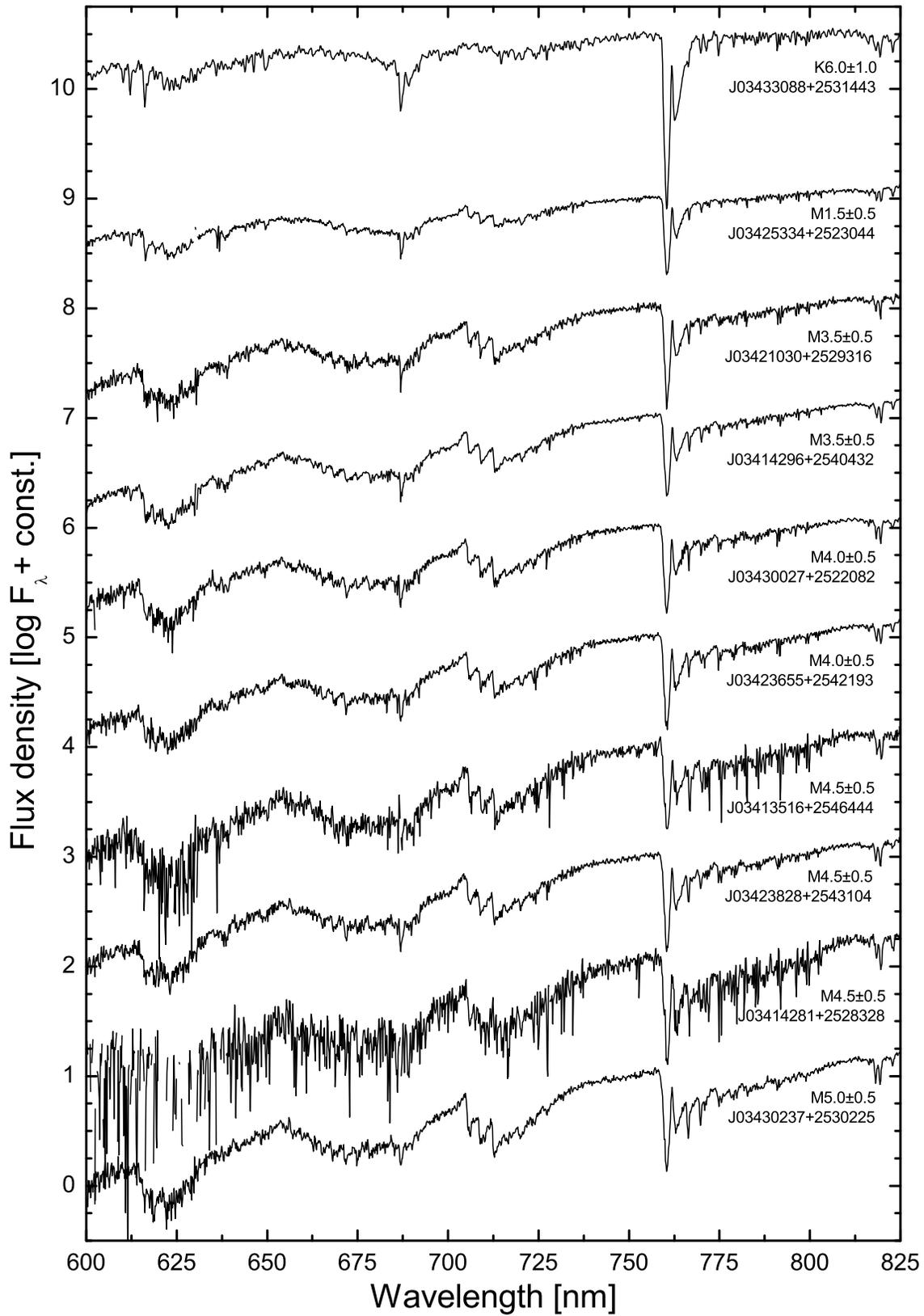}
\caption{FORS spectra of the ten objects sorted by their spectral type. The spectra shown here are not extinction corrected.}
\label{fig:FORS_spectra1}
\end{figure*}

\begin{figure*}[htb]\centering
\includegraphics[width=0.9\textwidth]{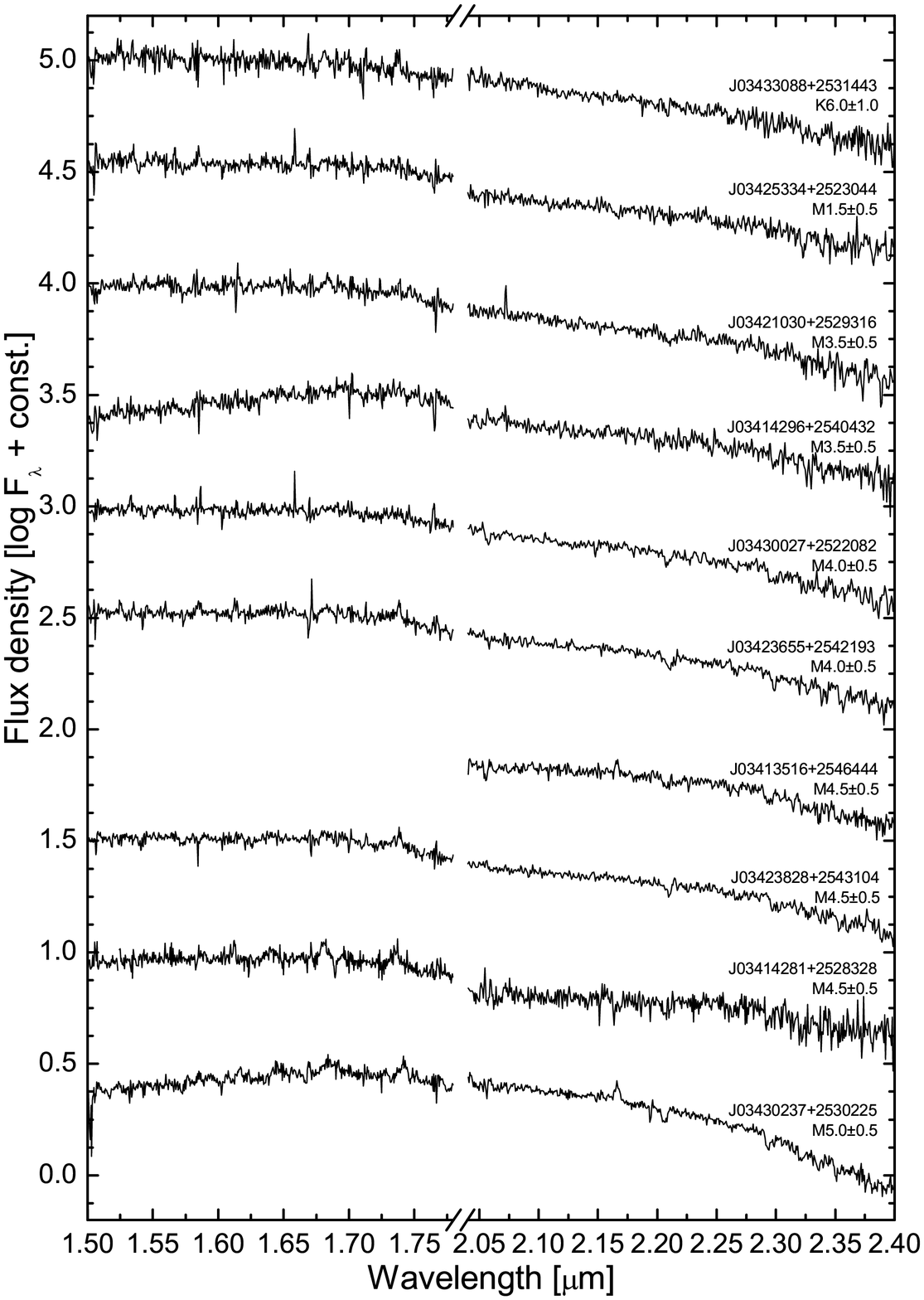}
\caption{ISAAC H and K band spectra of the ten objects sorted by their spectral type. A flux calibration between H and K band has not been performed. The different shapes of the H band spectrum of J03414296+2540432 and the K band spectrum of J03430237+2530225 are most probably due to problems with their standard star. For the H band spectrum of J03413516+2546444 no standard star was observed.}
\label{fig:ISAAC_spectra1}
\end{figure*}

The spectra obtained with FORS and ISAAC are shown in Fig.~\ref{fig:FORS_spectra1} and Fig.~\ref{fig:ISAAC_spectra1}. The spectra shown are not corrected for extinction. Almost all infrared spectra look similar. The different shapes of the ISAAC spectra of J03414296+2540432 and J03430237+2530225 are most probably due to calibration errors of the observed standard star, since there is no further hint of a different spectral type. No H band standard star was observed for J03413516+2546444.

\subsection{Spectral classification}

The previously described methods for spectral classification show that in all cases the actual spectral type is earlier than predicted by Eisenbeiss et al. (\cite{Eisenbeiss2009}). In two cases (J03433088+2531443 and J03425334+2523044) the spectral type was found to be much earlier (K6 and M1.5, respectively). In all the other cases a spectral type between M3.5 and M5 was determined (see Table~\ref{tab_SpT}). As expected, the TiO-5 ratio results in a quite accurate  determination of spectral type while the uncertainty for the VO spectral type is larger by a factor of two. Furthermore, VO does not provide any prediction for spectral types earlier than M4, since VO is not present in early M stars. For J03414281+2528328 and J03413516+2546444 the FORS-spectra turned out to be noisy around the TiO-5 feature leading to a larger error.

Overall, the determination of spectral type via the TiO-5 index and the comparison to template spectra are in good agreement. The fitting code of Sebastian \& Guenther (\cite{Sebastian2011}) led to similar results than the by-eye comparison with an extinction corrected spectrum. The best-fitting extinction derived by the code was on average $A_V\approx 0.6-1.3\,$mag and thus slightly higher but in agreement with the assumption. Since there are less sharp features present in M-type stars compared to A-type stars for which the program was originally designed, the error bars of spectral classification are larger, but prove the usability of the code even at the late end of the spectral sequence.

\begin{table*}[htb]
	\centering
	\caption{Results of the spectral analysis. Column two and three contain the spectral type and extinction from Eisenbeiss et al. \cite{Eisenbeiss2009}. The following columns contain the results of this study. The last column denotes the $A_V$ obtained from the five best-fitting spectral types from the fitting code of Sebastian \& Guenther (\cite{Sebastian2011}).}
	\label{tab_SpT}
	\begin{tabular}{l*3{r@{$\,...\,$}l}cccr@{$\,...\,$}l}
		\hline
		2MASS-number      & \multicolumn{4}{c}{Eisenbeiss et al. \cite{Eisenbeiss2009}}  & \multicolumn{4}{c}{visual spectrum} & final& \multicolumn{2}{c}{$A_V$}\\               
			          & \multicolumn{2}{c}{spectral type}&\multicolumn{2}{c}{$A_V [\textnormal{mag}]$}&\multicolumn{2}{c}{overall shape} & TiO-5& VO   & spectral type & \multicolumn{2}{c}{[mag]}\\  % 
		\hline
		J03433088+2531443 & M4&M6    &0.00&1.33 & K5&M0   & K6.4$\,\pm\,$0.5 & $\,\leq\,$M4     & K6.0$\,\pm\,$1.0 & 0.8& 1.3\\            
		J03430237+2530225 & M5&M6.5  &0.03&1.41 & M4.5&M6 & M4.8$\,\pm\,$0.5 & M4.7$\,\pm\,$1.6 & M5.0$\,\pm\,$0.5 & 1.0& 1.2\\            
		J03430027+2522082 & M6&M7    &0.08&1.53 & M3&M4.5 & M3.6$\,\pm\,$0.6 & M4.4$\,\pm\,$1.7 & M4.0$\,\pm\,$0.5 & 0.3& 0.8\\            
		J03425334+2523044 & M6.5&M7.5&0.05&1.45 & M1&M2.5 & M0.9$\,\pm\,$0.6 & $\,\leq\,$M4     & M1.5$\,\pm\,$0.5 & 0.9& 2.2\\            
		J03423828+2543104 & M6.5&M7.5&0.00&0.58 & M4&M6   & M4.0$\,\pm\,$0.6 & M4.5$\,\pm\,$1.4 & M4.5$\,\pm\,$0.5 & 0.8& 1.8\\            
		J03423655+2542193 & M6&M7    &0.04&1.24 & M3.5&M5 & M3.5$\,\pm\,$0.6 & M4.1$\,\pm\,$1.6 & M4.0$\,\pm\,$0.5 & 1.3& 1.5\\            
		J03421030+2529316 & M5&M7    &0.01&1.40 & M3&M4.5 & M3.3$\,\pm\,$0.6 & M3.5$\,\pm\,$1.4 & M3.5$\,\pm\,$0.5 & 0.5& 1.3\\            
		J03414296+2540432 & M7.5&M8  &0.00&1.34 & M3&M4.5 & M3.0$\,\pm\,$0.8 & M4.6$\,\pm\,$1.1 & M3.5$\,\pm\,$0.5 & 0.7& 1.6\\            
		J03414281+2528328 & M5&M6    &0.02&1.44 & M4.5&M6 & M3.3$\,\pm\,$1.4 & M4.3$\,\pm\,$3.5 & M4.5$\,\pm\,$1.0 & 0.6& 1.7\\            
		J03413516+2546444 & M5&M6    &0.00&1.24 & M4&M5.5 & M3.9$\,\pm\,$1.1 & M4.5$\,\pm\,$1.8 & M4.5$\,\pm\,$1.0 & 0.9& 1.2\\            
		\hline
	\end{tabular} 
\end{table*}

\subsection{Luminosity classes}

\begin{figure}[htb]
\includegraphics[width=1\columnwidth]{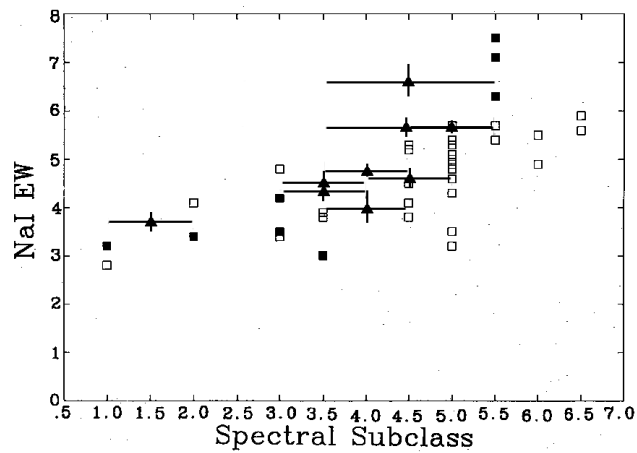}
\caption{The relation between sodium equivalent width and spectral subclass for M-type dwarfs. Field dwarfs are marked with filled squares. Pleiades dwarfs marked with open squares (from Steele \& Jameson \cite{Steele1995}; their Fig.~8). The objects of this study are marked by filled triangles. Object J03433088+2531443 is not plotted. For further explanations see text.}
\label{fig:Steele_sodium}
\end{figure}

As described above, the luminosity class can be determined using several methods. Steele \& Jameson (\cite{Steele1995}) investigated the equivalent width of the sodium doublet at $\lambda=8183/8195\,$\AA{}  for field dwarfs and dwarfs in the Pleiades. The location of the objects of this study in their plot is shown in Fig.~\ref{fig:Steele_sodium}. They clearly follow the relation for mid M-type dwarf stars. Hence, those nine stars from our sample are dwarf stars. The late K-type star J03433088+2531443 is not visible in the range of the plot. Extrapolating the trend seen in the figure, it might also be a dwarf, but since no late K-dwarfs are part of the study of Steele \& Jameson (\cite{Steele1995}) this only an estimate.

\begin{table*}[htb]
	\centering
	\caption{Determination of luminosity classes; TiO-5 and CaH-2 index, equivalent widths of sodium and potassium lines and line depth comparison between sodium and CO. According to these numbers, all objects are dwarf stars. For further explanation see text.}
	\label{tab_luminosityclass}
	\begin{tabular}{lcccccc}
		\hline
		2MASS-number      & TiO-5             & CaH-2 & \multicolumn{3}{c}{equivalent widths [\AA{}]} & Na \& Ca\\               % 
		                  &                   &      & \footnotesize{K\,I ($\lambda=7665\,$\AA{})}& \footnotesize{K\,I ($\lambda=7699\,$\AA{})} & \footnotesize{Na\,I ($\lambda=8183/8195\,$\AA{})} & similar CO\\  % 
		\hline                        
		J03433088+2531443 & 0.91$\,\pm\,$0.02 & 0.85$\,\pm\,$0.04 & 0.5$\,\pm\,$0.1 & 1.2$\,\pm\,$0.1 & 2.1$\,\pm\,$0.1 & (yes)\\            
		J03430237+2530225 & 0.31$\,\pm\,$0.01 & 0.37$\,\pm\,$0.03 & 4.3$\,\pm\,$0.2 & 2.3$\,\pm\,$0.1 & 5.6$\,\pm\,$0.1 &  yes \\            
		J03430027+2522082 & 0.43$\,\pm\,$0.03 & 0.45$\,\pm\,$0.03 & 1.2$\,\pm\,$0.1 & 0.9$\,\pm\,$0.1 & 4.0$\,\pm\,$0.3 &  yes \\            
		J03425334+2523044 & 0.68$\,\pm\,$0.04 & 0.59$\,\pm\,$0.02 & 1.0$\,\pm\,$0.1 & 0.8$\,\pm\,$0.1 & 3.8$\,\pm\,$0.2 &  --  \\            
		J03423828+2543104 & 0.41$\,\pm\,$0.02 & 0.38$\,\pm\,$0.04 & 2.4$\,\pm\,$0.1 & 1.2$\,\pm\,$0.1 & 4.5$\,\pm\,$0.2 &  yes \\            
		J03423655+2542193 & 0.44$\,\pm\,$0.03 & 0.42$\,\pm\,$0.04 & 2.2$\,\pm\,$0.1 & 1.5$\,\pm\,$0.3 & 4.6$\,\pm\,$0.1 &  yes \\            
		J03421030+2529316 & 0.46$\,\pm\,$0.03 & 0.44$\,\pm\,$0.03 & 1.4$\,\pm\,$0.1 & 1.1$\,\pm\,$0.1 & 4.2$\,\pm\,$0.2 &  yes \\            
		J03414296+2540432 & 0.48$\,\pm\,$0.06 & 0.43$\,\pm\,$0.02 & 2.0$\,\pm\,$0.1 & 1.4$\,\pm\,$0.1 & 4.4$\,\pm\,$0.2 &   -- \\            
		J03414281+2528328 & 0.46$\,\pm\,$0.12 & 0.39$\,\pm\,$0.13 & 2.8$\,\pm\,$0.1 & 1.8$\,\pm\,$0.1 & 6.7$\,\pm\,$0.3 &  yes \\            
		J03413516+2546444 & 0.40$\,\pm\,$0.09 & 0.35$\,\pm\,$0.06 & 2.9$\,\pm\,$0.1 & 2.5$\,\pm\,$0.1 & 5.6$\,\pm\,$0.2 & (yes)\\            
		\hline                                                       
	\end{tabular}                                                 
\end{table*}                                                   

Using the equivalent width of the sodium doublet at $\lambda=5890/5896\,$\AA{} as an indicator as proposed by Hamilton \& Stauffer (\cite{Hamilton1993}) turned out not to be useful as the spectra have been to noisy around this doublet. Following Hamilton \& Stauffer (\cite{Hamilton1993}), a second feature, which is indicative for dwarf nature is the K\,I resonance doublet at $\lambda=7665/7699\,$\AA{}. Unfortunately they do not give any reference values for their field dwarfs, but only write that the strength of this feature is an indicator for luminosity. Apart from J03433088+2531443 and J03425334+2523044, e.g. the two early type stars, and J03430237+2530225, e.g. the latest star, all the others show similar equivalent widths for $\lambda=7665\,$\AA{}. Regarding the feature at $\lambda=7699\,$\AA{} all stars show similar values. The discrepancy in the first case might be due to the spectral class. Hamilton \& Stauffer (\cite{Hamilton1993}) also mention a temperature dependence of the K I feature \textit{``due to the low ionization potential for potassium and because that they are resonance lines.''}

\begin{figure}[htb]
\includegraphics[width=1\columnwidth]{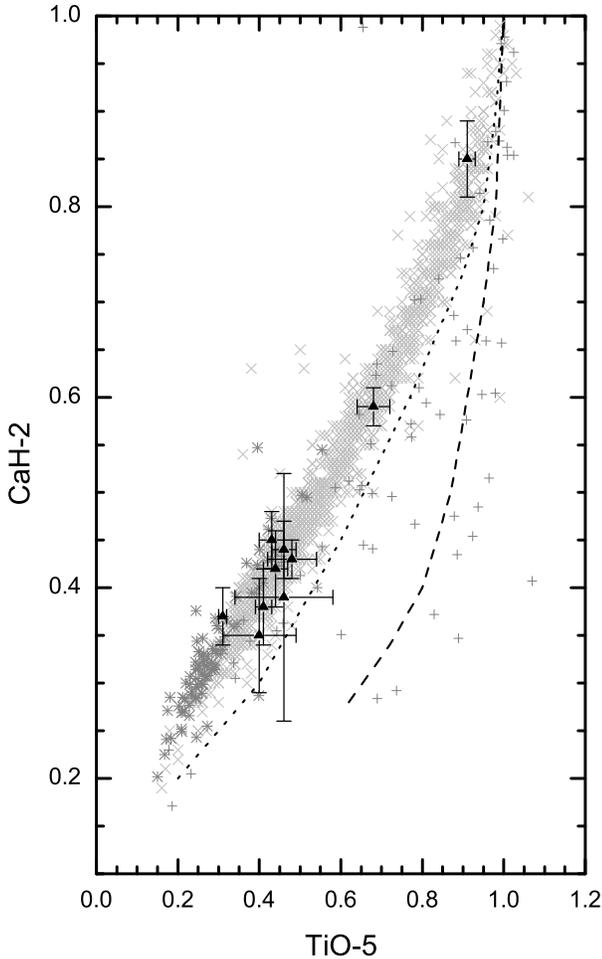}
\caption{The relation between CaH-2 and TiO-5 for late K-type to late M-type stars. Reference data taken from Kirkpatrick et. al \cite{Kirkpatrick1999} (crosses), Mart\'in et al. \cite{Martin1999} (plus signs) and Cruz \& Reid \cite{Cruz2002} (asterisks). Black triangles denote the data from the objects of this study according Table~\ref{tab_luminosityclass}. The dotted and dashed lines mark the approximate border between dwarfs and subdwarfs, and subdwarfs and extreme subdwarfs, respectively.}
\label{fig:CaH-2_TiO5}
\end{figure}

Plotting the objects of this study into the TiO-5--CaH-2 diagram (see Fig.~\ref{fig:CaH-2_TiO5}) analogously to Cruz \& Reid (\cite{Cruz2002}) one can clearly see that none of them coincides with the area of M-type subdwarfs (sdM) or extreme subdwarfs (esdM), hence none of them shows enhanced metalicity. Only the stars J03414281+2528328 and J03413516+2546444 reach the sub-dwarf area due to their big error bars, but they are expected not to be sdM or esdM stars as well.

Finally, we considered the molecular and atomic absorption lines in the K band. Here, the CO band heads at $\lambda\approx2.28-2.38\,\mu$m are very prominent features for M-type stars. As described in Section~\ref{sec:spectralindices}, especially the sodium doublet at $\lambda=2.21\,\mu$m is comparably strong in dwarf stars as well as the calcium triplet at $\lambda=2.26\,\mu$m. By comparing the strength of these lines and band heads one can determine the luminosity class. As listed in Table~\ref{tab_luminosityclass} this similarity can be clearly identified in six objects. In two objects (J03433088+2531443 and J03413516+2546444) this can be seen as well, although both structures are quite weak and noisy. In the remaining two objects (J03425334+2523044 \& J03414296+2540432) not even the CO bandheads can be seen clearly, so that no judgement is possible here.

From all luminosity class indicators together, it can be concluded that all observed objects are dwarf stars.

\subsection{Lithium absorption and H$\alpha$ emission}

All Pleiades members with $M_I>7.5\,$mag show H$\alpha$ emission (Hodgkin, Jameson \& Steele \cite{Hodgkin1995}) due to chromospheric activity. In three of our stars there is no H$\alpha$ emission present. In five stars a very low upper limit of up to $W_{\textnormal{H}\alpha}<0.5\,$\AA{} can be given. All these cases can be treated as nondetections and therefore imply non-membership. 
Two objects (J03430237+2530225 and J03423828+2543104) show H$\alpha$ emission with an equi-valent width of $W_{\textnormal{H}\alpha} = 1.5\,$\AA{} and $W_{\textnormal{H}\alpha} = 1.1\,$\AA{}, respectively, which would still be below the lower end of the values given in Hodgkin et al. (\cite{Hodgkin1995}) for Pleiades members.

A second study of the H$\alpha$ emission as a membership indicator was done by Prosser, Stauffer \& Kraft (\cite{Prosser1991}). On comparison with their results, again the two stars from our study with the largest emission lines show a value too low for being classified as Pleiades members with mid-M spectral type. It should be noted though, that the already mentioned study from Hodgkin et al. (\cite{Hodgkin1995}) did measure some late type Pleiades members with lower H$\alpha$ emission than the Prosser et al. (\cite{Prosser1991}) study.

Following the conclusion of Hamilton \& Stauffer (\cite{Hamilton1993}) these measurements would imply that the stars observed in our study are not members of the Pleiades.

\begin{table}[htb]
	\centering
	\caption{Equivalent widths of lithium absorption ($\lambda=6708\,$\AA{}) and H$\alpha$ emission ($\lambda=6563\,$\AA{}).}
	\label{tab_lithium_halpha}
	\begin{tabular}{lccccc}
		\hline
		2MASS-number      & \multicolumn{2}{c}{equivalent widths [\AA{}]}\\               % 
		                  &  Li absorption &  H$\alpha$ emission \\  % 
		\hline                        
		J03433088+2531443 & 0       & 0               \\            
		J03430237+2530225 & $<$0.12 & 1.5$\,\pm\,$0.6 \\            
		J03430027+2522082 & 0       & $<$0.5          \\            
		J03425334+2523044 &  0      & 0               \\            
		J03423828+2543104 & 0       & 1.1$\,\pm\,$0.4 \\            
		J03423655+2542193 &  0      & $<$0.3          \\            
		J03421030+2529316 & 0       & $<$0.5          \\            
		J03414296+2540432 & $<$0.03 & $<$0.2          \\            
		J03414281+2528328 &  0      &  0              \\            
		J03413516+2546444 & $<$0.25 & $<$0.5          \\            
		\hline                                                       
	\end{tabular}                                                 
\end{table}   

At an age of 119\,Myr (Mart\'in, Dahm \& Pavlenko \cite{Martin2001}) all the lithium in the Pleiades stars should be depleted by at least a factor of 1000 (Rebolo, Mart\'in \& Maguzz\`u \cite{Rebolo1992}). Taking into account that our stars are older, it is not expected to find any lithium at all unless they are brown dwarfs. As noted by Rebolo et al. (\cite{Rebolo1992}), \textit{''lithium should be preserved in substellar objects.``} As shown in Table~\ref{tab_lithium_halpha} no lithium is present in most objects. For three of them very low lithium absorption can be measured, too low to be brown dwarfs. 

According to Steele \& Jameson (\cite{Steele1995}) the upper limit for lithium absorption in the Pleiades is $W_{\textnormal{Li}}\sim0.2\,$\AA{}, which is of the order of magnitude of the equivalent width limits in J03413516+2546444 and J03430237+2530225. Using lithium as a youth indicator it can be concluded that all the objects are at least as old as the Pleiades, but most probably older.

\subsection{Distance and proper motion}\label{sec_dist_propmotion}

Deriving the distance to our objects via the distance modulus requires the knowledge of absolute and apparent magnitudes as well as the extinction. Reliable measurements of the apparent magnitude have been made by 2MASS where JHK-magnitudes are available. As listed in equations\,\ref{eqn:absmag1}--\ref{eqn:absmag6} Cruz \& Reid (\cite{Cruz2002}) determined a relation between the TiO-5, CaH-2 and CaOH index and the absolute magnitude in J band for M-type dwarfs. If the valid equations are distinct, the weighted mean of the three magnitudes has been calculated. In the cases where the absolute magnitude is not uniquely given by one equation the weighted mean of the upper and lower value has been calculated. Finally, the mean has been used as magnitude and the maximum absolute deviation as magnitude error. With the extinction relation $A_V/A_J=0.282$ from Rieke \& Lebofsky (\cite{Rieke1985}) one can take an average extinction of $A_V=0.75\,$mag (see Eisenbeiss et al. \cite{Eisenbeiss2009}) as a correction term remembering that this might underestimate the true extinction and therefore overestimates the distance. But since extinction is very low in the infrared the effect on the spectrophotometric distances is not very significant. The resulting distances are shown in Table~\ref{tab_distance}.

\begin{table}[htb]
	\centering
	\caption{Absolute magnitude $M_J$ from equations\,\ref{eqn:absmag1}--\ref{eqn:absmag6}, apparent magnitudes $m_J$ from 2MASS and the extinction corrected distance to the ten objects.}
	\label{tab_distance}
        \footnotesize
	\begin{tabular}{lccr@{$\,\pm\,$}r}
		\hline
		2MASS-number      & $M_J$             & $m_J$              & \multicolumn{2}{c}{$d$}\\               % 
		                  &  [mag]            & [mag]              & \multicolumn{2}{c}{[pc]}\\  % 
		\hline                                                      
		J03433088+2531443 & 5.69$\,\pm\,$0.20 & 17.01$\,\pm\,$0.20 & 1662 & 304 \\          
		J03430237+2530225 & 8.21$\,\pm\,$0.24 & 15.04$\,\pm\,$0.04 &  211 &  27 \\         
		J03430027+2522082 & 8.16$\,\pm\,$0.71 & 16.44$\,\pm\,$0.11 &  412 & 156 \\         
		J03425334+2523044 & 6.59$\,\pm\,$0.21 & 16.47$\,\pm\,$0.13 &  858 & 133 \\         
		J03423828+2543104 & 8.43$\,\pm\,$0.70 & 15.52$\,\pm\,$0.05 &  238 &  82 \\         
		J03423655+2542193 & 8.14$\,\pm\,$0.72 & 15.84$\,\pm\,$0.08 &  315 & 115 \\         
		J03421030+2529316 & 7.49$\,\pm\,$0.23 & 16.42$\,\pm\,$0.11 &  553 &  88 \\         
		J03414296+2540432 & 8.10$\,\pm\,$0.69 & 16.16$\,\pm\,$0.10 &  371 & 134 \\         
		J03414281+2528328 & 8.17$\,\pm\,$1.13 & 15.95$\,\pm\,$0.08 &  327 & 183 \\         
		J03413516+2546444 & 8.64$\,\pm\,$0.92 & 16.38$\,\pm\,$0.11 &  321 & 153 \\         
		\hline                                                       
	\end{tabular}                                                 
\end{table}   

Compared to the Pleiades distances from parallax measurements from HIPPARCOS ($d=120.2\,$pc; van Leeuwen \cite{vanLeeuwen2009}) and from main sequence fitting ($d=135.5\,$pc; An et al. \cite{An2007}) all our objects are behind the Pleiades within the respective error bars. As shown in Table~\ref{tab_SpT} the extinction values derived by the fitting code of Sebastian \& Guenther (\cite{Sebastian2011}) are slightly higher than the average extinction in this region used as an assumption for deriving the distance, also in agreement with the background hypothesis.

\begin{figure}[htb]
\includegraphics[width=1\columnwidth]{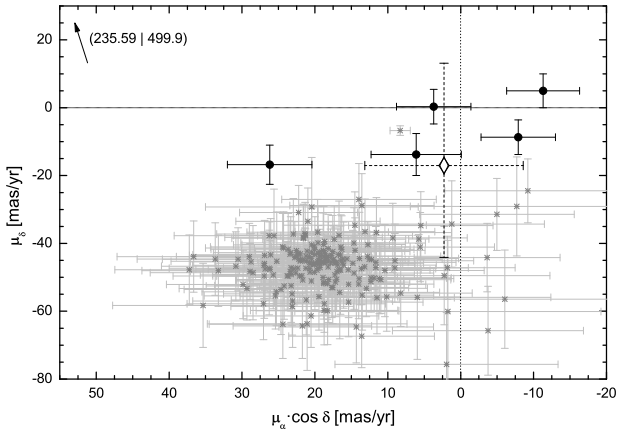}
\caption{The figure shows the proper motion (in $\textnormal{RA}\cdot\cos(\textnormal{DEC})$ and DEC) of Pleiades members (gray asterisks) and the 6 objects of this study with known proper motion (black dots). Only Pleiades members with a membership probability $>$75\% according to Deacon \& Hambly (\cite{Deacon2004}) and Schilbach et al. (\cite{Schilbach1995}) were taken into account; proper motions come from the PPMXL catalog (Roeser et. al \cite{Roeser2010}) and in case of J03430027+2522082, upper left, from the NOMAD catalog (Zacharias et al. \cite{Zacharias2004}). The diamond represents our proper motion analysis result of J03430027+2522082, see Table \ref{tab_posFastStar} and text in Sect.~\ref{sec_dist_propmotion}, not consistent with the NOMAD value, but with zero.}
\label{fig:propermotion_mu}
\end{figure}

\begin{table}[htb]
	\centering
	\caption{Proper motions of six objects of our study from the PPMXL catalog (Roeser et. al \cite{Roeser2010}; the upper five stars) and NOMAD (Zacharias et al. \cite{Zacharias2004}; the sixth star) together with the mean proper motion of the Pleiades according to Robichon et al. (\cite{Robichon1999}).}
	\label{tab_propermotion}
	\begin{tabular}{l*2{r@{$\,\pm\,$}l}}
		\hline
		2MASS-number      & $\mu_{\alpha}\cos\delta$&$\sigma_{\mu_{\alpha}\cos\delta}$ & $\mu_{\delta}$&$\sigma_{\mu_{\delta}}$ \\               % 
		                  &  \multicolumn{2}{c}{[mas\,yr$^{-1}$]}            & \multicolumn{2}{c}{[mas\,yr$^{-1}$]}\\  % 
		\hline                                                      
		J03433088+2531443 & 3.7   & 5.1  & 0.3    & 5.1\\          
		J03430237+2530225 & -11.3 & 5.0  & -33.4  & 5.0\\         
		J03425334+2523044 & 6.1   & 6.2  & -13.8  & 6.2\\         
		J03423828+2543104 & -7.9  & 5.1  & -8.7   & 5.1\\         
		J03421030+2529316 & 26.2  & 5.8  & -16.8  & 5.8\\ 
		\hline
		J03430027+2522082 & 235.9 & 9.0  & 499.9  & 9.0\\         
		\hline                                              
		Pleiades center   & 19.15 & 0.23 & -45.72 & 0.18 \\ 
		\hline                                              
	\end{tabular}                                                 
\end{table} 
In the light of these results we expect our stars not to share the Pleiades mean proper motion. As shown in Fig.~\ref{fig:propermotion_mu}, all six objects for which proper motion measurements are available fulfill this expectation. The respective values are given in Table~\ref{tab_propermotion}. As can be seen, four of the six objects are consistent with zero proper motion (within $2\,\sigma$) and hence with being background objects.

Object J03430027+2522082 shows a very large proper motion, based on NOMAD catalog data. Since NOMAD only refers to older catalogs, a redetermination of the proper motion is advisable. Therefore we retrieved images of the ESO Second Digitalized Sky Survey (Lasker \& STSCI Sky-Survey Team \cite{Lasker1998}; DSS-2), and the UKIRT Infrared Deep Sky Survey (Lawrence et. al \cite{Lawrence2007}; UKIDSS). The observations of the particular region of interest are from 1989 December 17 and 2005 October 8, respectively. A new astrometric calibration was performed using \textit{GAIA}\footnote{Graphical Astronomy and Image Analysis Tool; available at http://astro.dur.ac.uk/$\sim$pdraper/gaia/gaia.html}. The position was measured using the \textit{Source Extractor} (Bertin \& Arnouts \cite{Bertin1996}). As shown in Table~\ref{tab_posFastStar} the differences in the positions do not correspond to the proper motion given in NOMAD, where a movement of many arcseconds would have been expected. Our measurements are consistent with no proper motion ($\mu_{\alpha}\cos\delta=2.28^{+10.88}_{-10.86}\:$mas/yr; $\mu_{\delta}=-17.0^{+30.1}_{-27.0}\:$mas/yr) within the respective error bars. The large proper motion value in NOMAD could have resulted from a mis-identification in one of the three epochs.

\begin{table}[htb]
	\centering
	\caption{Position measurements of J03430027+2522082. The images used for the measurements have been taken from ESO DSS-2 (red plate) and the UKIDSS (Y-filter) with a epoch difference of 15.81 years.}
	\label{tab_posFastStar}
	\begin{tabular}{lrr}
		\hline
		image source  & ESO DSS-2 & UKIDSS\\
		\hline
		date of observation & 1989 Dec 17 & 2005 Oct 08\\
		$\alpha$ [hh:mm:s.ss] & 03:43:0.28 & 03:43:0.32\\
		$\Delta\alpha$ [s] & $\pm$0.11 & $\pm$0.08\\
		$\delta$ [dd:mm:s.ss] & +25:22:8.35 & +25:22:8.08\\
		$\Delta\delta$ [s] & $\pm$0.67 & $\pm$0.39\\
		\hline                                              
	\end{tabular}                                                 
\end{table}

\section{Conclusion}

In this paper we presented the investigation of ten brown dwarf candidates in the Pleiades discovered by Eisenbeiss et al. (\cite{Eisenbeiss2009}). The analysis of spectral type and luminosity class showed that all objects are dwarf stars earlier than M5.5 that show no evidence of enhanced lithium abundances and therefore are neither brown dwarfs nor very young objects. The spectrophotometric distances show, that all stars are behind the Pleiades. Furthermore, six objects for which proper motion values are available can be kinematically excluded as Pleiades members.
Two stars show small H$\alpha$ emission probably indicating chromospheric activity typical for M-type stars.

The discrepancies between the results of this work and those of Eisenbeiss et al. (\cite{Eisenbeiss2009}) show the difficulty of deriving spectral types from photometric data, especially for M-type stars. The simultanous fitting of extinction is no trivial task and gives only a first guess of the spectral type that has to be confirmed using spectra. 
Furthermore, especially the R band exposures have been made at low altitudes and hence high airmass. In addition, known Pleiades members have been used as photometric calibrators. Since they are young, they might be variable and hence led to calibration errors. Finally, before adding up, all images have been corrected for field distortions to apply better astrometric precision. Both issues might have lead to underestimated brightness and hence, later spectral types.

%%%
%%% -MWL- 2006-01-13 auf Verlagswunsch wieder altes Bibliographie-Format
%%% 

\acknowledgements MS and RN acknowledge general support from the German National Science Foundation (Deutsche Forschungsgemeinschaft, DFG) in grant NE 515/36-1. TE would like to thank DFG for support in project NE 515/30-1 and in SFB-TR 7. MS would like to thank all members of the AIU Jena for their help and support during diploma phase.

Based on observations made with ESO telescopes at the La Silla Paranal Observatory under programme ID 084.C-0654. 
This research has made use of the VizieR catalog access tool, operated at the Observatoire Strasbourg, as well as of the WEBDA database, operated at the Institute for Astronomy of the University of Vienna. This research has benefited from the M dwarf standard spectra made available by Kelle Cruz.

\newpage%%%%%%%%%%%%%%%%%%%%%%%%%%%%%%%%%%%%%%%%%%%%%%%%%%%%%%

\end{document}